\documentclass[conference]{IEEEtran}
\IEEEoverridecommandlockouts
\usepackage{cite}
\usepackage{amsmath,amssymb,amsfonts}
\usepackage{algorithmic}
\usepackage{graphicx}
\usepackage{textcomp}
\usepackage{xcolor}
\def\BibTeX{{\rm B\kern-.05em{\sc i\kern-.025em b}\kern-.08em
    T\kern-.1667em\lower.7ex\hbox{E}\kern-.125emX}}
\begin{document}

\title{IEC61850 Sample-Value Service Based on Reduced Application Service Data Unit for Energy IOT\\
{\footnotesize \textsuperscript{*}}
\thanks{Identify applicable funding agency here. If none, delete this.}
}

\author{\IEEEauthorblockN{1\textsuperscript{st} Wenhao Xu}
\IEEEauthorblockA{\textit{College of Electric Engineering and Automation} \\
\textit{Fuzhou University}\\
Fuzhou, China \\
1281505942@qq.com}
\and

\IEEEauthorblockN{2\textsuperscript{th} Nan Xie}
\IEEEauthorblockA{\textit{College of Electric Engineering and Automation} \\
\textit{Fuzhou University}\\
Fuzhou, China \\
t13056@fzu.edu.cn(Corresponding Auther)}
}

\maketitle

\begin{abstract}
With the development of 5G technology and low-power wireless communication technology, a large number of IOT devices are introduced into energy systems. Existing IOT communication protocols such as MQQT and COAP cannot meet the requirements of high reliability and real-time performance. However, the 61850-9-2 Sample value protocol is relatively complex and the message length is large, difficult to ensure real-time transmission for IOT devices with limited transmission rate. This paper proposes a 9-2 SV protocol for IOT controller based on Application Service Data Unit. The communication protocol is strictly in accordance with IEC61850-9-2 and can be recognized by existing intelligent electronic devices such as merging units. The protocol simplifies and trims some parameters, and changes the floating point value to integer data. Considering the instability of wireless communication, unicast or multicast UDP/IP is utilized to send SV Payload based on the 2.4GHz WIFI. The maximum transmission rate can be up to 30 Mbps. The hardware to implement reduced SV adopts ESP32-S, which is a dual-core MCU supporting WIFI with frequency of 240MHz. Software is based on FreeRTOS, LWIP and Libiec61850. A PC or raspberry PI is used as the Host to receive and analyze packets, verify feasibility of reduced SV protocols.
\end{abstract}


\section{Introduction}
As the development of Energy Internet and Smart Grid, the number of accessed distributed energy is continuously increasing. As result, It is urgent to introduce the technology of Internet of Thing (IOT)\cite{ref1}, Big Data and Artificial Intelligent for realization of wide-area time-space measurement and parameter identification, ensuring stability and maintainability of the whole energy system.

In recent years, wireless communication IOT technology have developed rapidly. Representative wireless-communication hardware technology, corresponding to the physical layer and link layer in communication model,  include WIFI, Bluetooth, Lora, NB-IOT, ZigBee \cite{ref2} and so on. The main communication protocols of the IOT, corresponding to the application layer or the network layer in the communication model, contains MQTT, COAP, TCP/IP, UDP/IP, Thread and so on.

The popular IOT communication protocols have shortcomings in reliability, stability and real-time performance.
For real-time consideration, the Goose and SV of IEC61850\cite{ref3,ref4} are usually adopted for critical switch operation and measurement of primary voltage and current in smart substation. These protocols internally support object-oriented characteristics. When transmitting four-teleportation signals based on digital and analog quantities, the meaning of transmission value is also given in the message, which is convenient for updating and maintenance of local database. While the advanced object-oriented method is not supported in the mainstream IOT protocols.

However, the 9-2 protocol is too flexible and complex, even the 9-2 LE still has a great burden on the transmission with IOT devices. These limitations are mainly reflected in the following aspects.

\paragraph{High sample rate}
9-2 LE stipulates that the sampling points of a one period are 80 and 256. For 50Hz power frequency signals,  80 sample points is acceptable to slow-rate IOT devices, while 256 sample points in one period is difficult to achieve at the present stage.

\paragraph{ Including float data }

Sampling values transmitted by SV are usually stored in floating-point format. For IOT devices with limited computing power, their storage and data post-processing bring great burden, which can only be better realized by high-speed DSP . However, pure hardware circuits such as FPGA and CPLD have inherent disadvantages in dealing with floating point arithmetic.

\paragraph{Too many useless Data objects}
The most important point is that there are too many data set parameters in the message, many of which are unnecessary for IOT devices and non-contact measurement. Therefore, the message length can be greatly cut down.

\begin{figure}[htbp]
	
	\centerline{\includegraphics[width=0.45\textwidth]{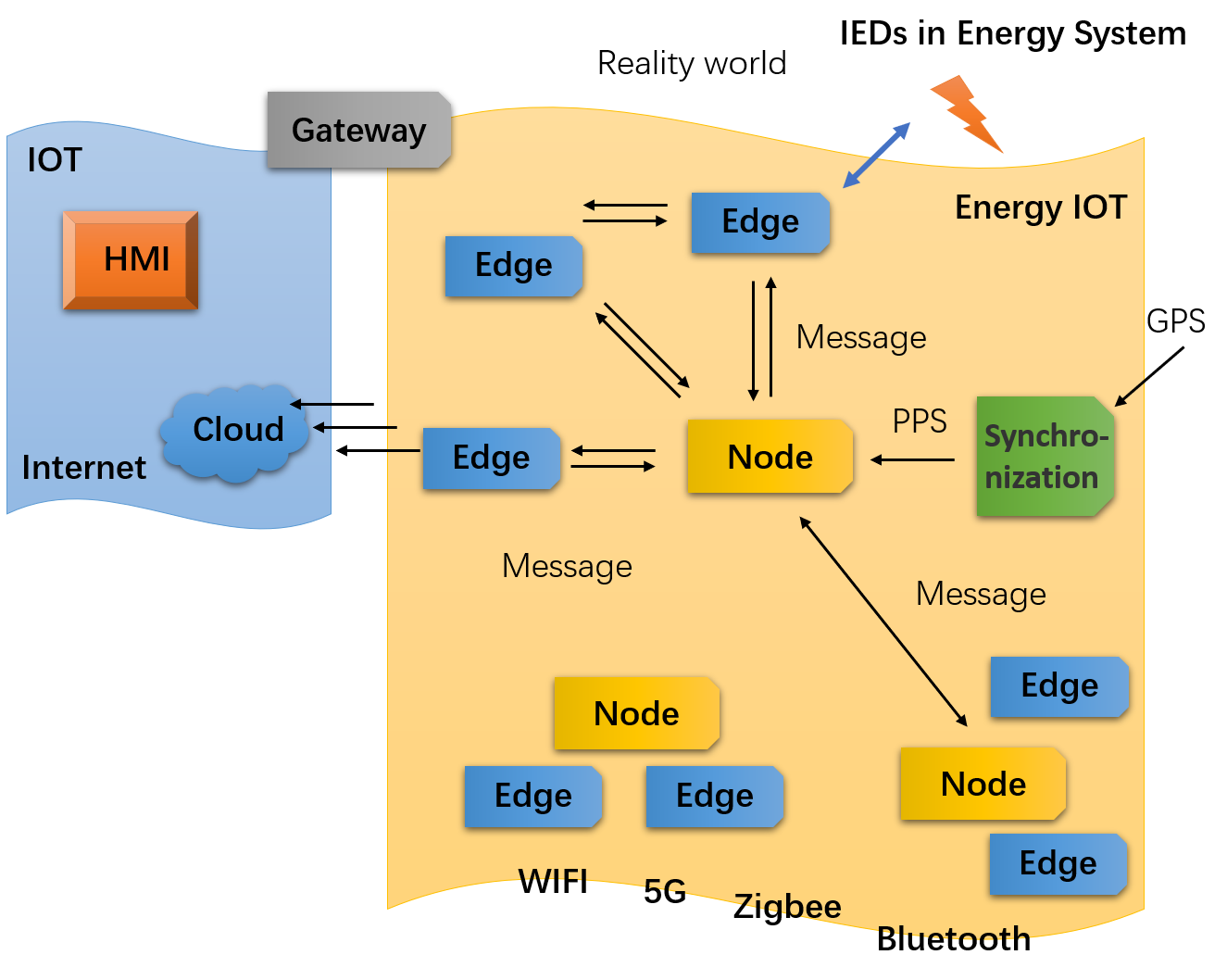}}
	\caption{The architecture of Energy IOT.}
	\label{fig}
\end{figure}

\paragraph{Lack of important Data objects}
Key objects of IOT applications, such as GPS spatial information, are not defined. For non-contact measurement applications, only magnetic field Logic Nodes and Data Objects are defined, while that for electric field is not.

Based on the above analysis, this paper proposes a reduced Applied Service Data Unit(ASDU) IEC61850-9-2 for Energy system served for IOT technology. The scheme can greatly reduce the length of message bytes by cutting redundant data sets and make them short enough for sending and receiving IOT messages. This scheme be compatible with the 9-2 standard. Messages based on UDP unicast/multicast technology are sent over WIFI2.4g. According to hardware corpoaration's private protocol, wifi signal can be transmitted in 1 km open area. The protocol has been implemented on IOT devices. The open source protocol library Libice61850 is used to create and update the packets, which is based on freeRTOS real-time operating system. The communication packets are detected and analyzed by Wireshark, which verifies the feasibility of the scheme.

\section{AIM and Energy IOT}
The aim of developing reduced SV is to provider a reliable communication protocol with real time performance for Energy IOT. The Energy IOT means the Energy systme plus IOT technology, a derivative of Energy Internet. Its focus is the interaction of things in power system such as Switch, Transducer, Transformer, sensor, distributed energy devices and so on. Its architecture
is shown in Fig 1. There are three key elements in the IOT Energy, Edge, Node and Gateway.

\begin{table}[htbp]
	\caption{Popular Wireless communication technology}
	\begin{center}
		\begin{tabular}{|p{1cm}|p{1.5cm}|p{1.5cm}|p{1cm}|p{1.5cm}|}
			
			\hline
			\textbf{Protocol}&\textbf{Chip}&\textbf{Transmission}&\textbf{Distance}&\textbf{Band}\\
			\hline 
			
			Blootooth 5.0 &	Espressif ESP32	& 4Mbps & 10m & 2.4GHz\\
			\hline
			
			WIFI 2.4GHz	& Espressif ESP32 &	150 Mbps & 50m &	2.4GHz\\
			\hline
			
			ZigBee &	Espressif ESP32-H2	& 250Kbps @2.4GHz &	10-100m &	2.4GHz, 915MHz, 868MHz\\
			\hline
			
			5G \newline NB-IOT &	EC616S	 & less than 100Kbps	& 2km-15km &	2500 MHz-3500 MHz\\
			\hline
			
			Lora &	SX1268 &	less than 300Kbps	& less than 5km &	410MHz-525MHz\\
			\hline

		\end{tabular}
		\label{tab1}
	\end{center}
\end{table}

\paragraph {Node}
Node is the most critical element in system with the highest priority and privilege. It has a group of computation unit, and maintain a local distributed database, maybe in the form of block-chain in the future. One node with the highest privillage called main node, which could be monitored and controlled by Human and Machine interface (HMI) and Cloud. Main node supervise other nodes, also has the function of recognizing, Authorizing, connecting and isolating other nodes. However, the relation of nodes could be absolutely equal also. And these nodes constitute a Multi-agent system. In general, the Node would not directly connected to he real physical devices, and It served as a distributed server with Frog computing capacity.

\paragraph{Edge}
Edge is also very important in system because of direct connection to physical device. Edge is usually connected to a sensor or switch, and could not be considered as a independent intelligent device(IED) in IEC61850. A group with at least one nodes and several edges can present the communication parts of IED. Usually, edges could be considered as the procedure layer in IEC61850, and Node represents the middle layer. Edge is often utilized a MCU or DSP with wireless connection. And their resource is often limit, can not afford a operation system 
such as linux. The tiny real-time system such as freeRTOS is operated on chip. To grantee the critical real-time operation, CPLD or even FPGA should be added to the edge module.

\paragraph{Gateway}
A gateway is also treated as a wireless router to maintain a local area network(LAN),distribute the IP and forward the message packet to certain device by IP NO. In energy IOT, Gateway is also a bridge between the Internet and Energy IOT,  mapping communication protocols between Energy IOT and extrnal Internet. 

In sum, the Energy IOT and Physical devices constitute a Cyber-physical System (CPS). With the HMI in local IOT or Internet, A supervisory control and data acquisition (SCADA) is also formed. There is no clear distinction between them, while CPS focus on the Cyber attribute, the safety issue should be considered at first. One important issue is the communication message, hackers would block it and replace  with their own, resulting in serious disaster. And this could be exclude by the block-chain technology.

However, this paper focus on reliability and real time performance, the safety would be considered in the future. However this paper is the base for future block-chain implement for Energy system.

\section{Analysis of protocols}
\subsection{Selection of physical layer protocols}
\paragraph{WIFI}
WIFI has almost replaced electrical cable and become the preferred communication method for personal computers, mobile phones and tablets. A new generation of WIFI technology based on 6G frequency has the maximum up and down transmission speed of 3.6 Gbps and 3 Gbps, which is close to the transmission limit of optical Fiber Ethernet with 10 Gbps. 

\paragraph{Bluetooth}
Compared with WIFI technology, the transmission speed of Bluetooth is much lower than WIFI. And the communication distance is also short, but the advantage is low power consumption. The latest Bluetooth 5.2 technology has been widely used in noise reduction ear-pods and other  consumer electronic application powered by battery.

\paragraph{Lora}
Lora technology aims to achieve long-distance transmission with low power consumption, but its transmission speed is limited. In addition, Lora technology is protected by patents and is not open source, which limits its development. 

\paragraph{NB-IOT}
NB-IOT, also known as Narrow Band IOT technology, is based on existing 4G or 5G mobile communication technology and is one of the most promising technologies in the future. Its communication distance can reach several kilometers. In the future, with the further development of 5G technology and the proposal of 6G technology, NB-IOT will further decrease its power consumption and increase transmission speed. 

\paragraph{Zigbee}
Zigbee based on IEEE 802.15.4 standards, has been widely promising since its invention, but due to the hardware cost and the lack of unified upper layer communication protocol, its development has been limited. The advantage of Zigbee is the extremely low power consumption, but the transmission bandwidth is not very high. In power system, ZigBee technology has been applied to non-contact measurement and achieved a sample rate with 660 point in one period \cite{ref2}.

To sum up, ZigBee, Lora, NBIOT and Bluetooth technologies have disadvantages such as slow transmission rate, limited communication distance and closed-source. WIFI is relatively mature and has the fastest transmission. However its power consumption is also highest due to the fast transmission. Thanks to the development of  low-power WIFI technology by using low-frequency clock and low transmission power, the WIFI power could be reduced dramatically.

\subsection{Review of IOT application  protocols and transmission protocols}
The existing IOT communication protocols have shortcomings in reliability, stability and real-time performance. The most popular up-layer protocols is MQTT/COAP based on TCP/UDP. As transmission layer protocols, TCP and UDP could also used for transmit message in IOT with aid of Internet technology such as JavaScript Object Notation(JSON) and so on. The following paragraphs would analyze them one by one.

\paragraph{MQTT}
MQTT is the most widely used protocol in the Internet and IOT, it has been widely used in mobile phones, tablets, wearable devices and other personal consumer electronics. MQTT is based on the publisher-subscriber model and TCP/IP protocol, so the transfer speed is not enough.

\paragraph{COAP}
COAP is also a common communication protocol in the Internet of Things. It is based on UDP/IP technology, does not guarantee the reliability of transmission, and has a fast speed. It support the publisher-subscriber model and send any payload.
COAP can replace MMS service and map ASCI services through JSON \cite{ref6}. It is expected to become the underlying communication protocol of IEC61850 in the future.

\paragraph{TCP/IP}
TCP/IP technology can only use a single point of connection, supporting the communication model of Server and Client, but with poor support for the popular publisher-subscriber model in IOT. This function can only be added through the upper layer protocol such as MQTT, but it will increase the packet length and the  transmission burden of IOT devices. 

\paragraph{UDP/IP}
UDP/IP technology can realize multicast function and support publisher-Subscriber model. In recent years, IPV6 based UDP multicast technology has been applied in the IOT, and its rapid development deserves attention. 

\paragraph{Thread}
Thread is an upper-layer network layer technology based on ZigBee, and is mainly used in intelligent household electrical appliances. Due to insufficient transmission speed, Thread may not be suitable for analog data sending that requires large bandwidth in power systems. However, it can be considered for transmission of low-speed meteorological parameters such as temperature and humidity. Since the Zigbee has been introduced into the non-contact measurement, Thread may also be studied in the future in Magnetic/Electrical field measurement.  

In sum, the popular IOT communication protocols need to be further developed to meet requirement of high reliability and transmission in Energy systems. The most promise one is the multi-cast technology based on UDP/IP and IPV6.

\subsection{Selection of IEC61850 service for IOT application}
\paragraph{MMS}
Although MMS has make great success in industrial automation, and has been successfully introduced into the power system, it still have some problems.Its structure is too complex and the encode/decode is very time consuming. Some group has develop a map from MMS into Coap\cite{ref6}. however, MMS can not replace Goose and SV because lack of real time performance

\begin{table}[htbp]
	\caption{Logic Nodes used in Energy IOT}
	\begin{center}
		\begin{tabular}{|c|c|p{2cm}|}
			\hline
			\textbf{Node name}&\textbf{Description}&\textbf{Data object for measurement}\\
			\hline
			\text{TMGF}&\text{Magnetic field sensor}&\text{MagFld[SAV]}\\
			\hline
			\text{TEEF}&\text{Electrical field sensor}&\text{EleFld[SAV]}\\
			\hline
			\text{TTMP}&\text{Temperature sensor Name}&\text{Tmp[SAV]}\\
			\hline
			\text{TVBR}&\text{Vibration sensor Name}&\text{Vbr[SAV]}\\
			\hline
			\text{THUM}&\text{Humidity sensor}&\text{Hmdt[SAV]}\\
			\hline
			\text{TCTR}&\text{Current transformer}&\text{AmpSv[SAV]}\\
			\hline
			\text{VCVR}&\text{Voltage transformer}&\text{VolSv[SAV]}\\
			\hline		
		\end{tabular}
		\label{tab1}
	\end{center}
\end{table}

\paragraph{Goose}
 Goose message is mainly used for operating critical switch in 
 intelligent substation, thus require extremely high stability and reliability. The use of unstable IOT technology  has great potential security risks. Also the Goose has only one ASDU, its extensibility is limit.

\paragraph{SV}
SV service is mainly used to transfer analog sampling values, at present for transmission of voltage and current instant value with high real-time performance, mainly used in electronic transducers. 

IOT has the unreliability of message transmission due to the unstable wireless transmission, whose transmission and reliability would be disturbed easily by the environment. Since the Goose and SV is lack of Internet Procotols(IP), that means without IP number, the publisher and subscriber is just identified by the Media Access Control Address (MAC). This would be a problem in IOT for device authorized and safety consideration. As a result, if Goose and SV is selected, they should be transmitted by UDP or TCP message with IP support. Considering the real time performance, UDP is more appropriate.

In addition to real-time performance, security and reliability are also under considerations with the highest priority. Since the Goose is often utilized for operating the critical switch, and the message is always transmitted in reliable optical fiber network. So it is unwise to using the IOT technology to transmit. And this paper did not consider the realization of GOOSE.

However, Sampled Value(SV) service just record the real time current or voltage value, the missing of several point value could be accepted. For critical application such as Protection. There are always four ADCs in parallel with two optical fiber output for protection. If the SV is send with wireless communication, the parallel communication-device number could be increased. The idea that replacing the high-cost device with cheap and unreliable civilian paralleled-connected devices has been verified in the aerospace area. With this idea, the SPACE X corporation has  succeed in decrease the cost dramatically without compromising reliability.

\section{Reduction of Sample Value protocols}
\subsection{Reduction of SV following LE edition}
Since the IEC61850-9-2 SV is too flexible. To make it more easier to be realized, Experts from ABB, 
Siemens and other famous Electrical corporation has introduce a simplified edition, called LE. The reduced SV for IOT is following the LE edition in many aspects, mainly including:

\begin{table}[htbp]
	\caption{Definition of SAV in Energy IOT}
	\begin{center}
		\begin{tabular}{|p{2cm}|p{1cm}|p{3.5cm}|p{0.5cm}|}
			\hline
			\textbf{Attribute Name}&\textbf{Attribute Type}&\textbf{Description and 				 Remarks}& \textbf{M/O /C}\\
			\hline
			instMag.i & INT32 & instantaneous Sampling value, integer, defined in formula & M \\
			\hline
			
			q & Quality & using only the two attributes:  validity and test & O/M \\
			\hline
			
			sVC.scaleFactor & INT8 & Sampling value scaling factor & O \\
			\hline
			
			sVc.offset & INT32 &  offset, set to 0 & O\\
			\hline
			
			GeoCrd.B & INT32 & Latitude of geographic coordinates; Range: -1.8E8 - +1.8E8; scaleFactor : -4; offset: 0, - and + represent N and S &  O\\
			\hline
			
			GeoCrd.L & INT32 & Longitude of geographic coordinates; Range:-9e7 - +9e7; scaleFactor: -4;
			Offset: 0, - and + represent E and W & O\\
			\hline
			
			GeoCrd.H & INT16 & ground height of Geographic coordinates; Range: -9999 to 9999; scaleFactor: -1; offset: 0 & O\\
			\hline  
			
			GeoCrd.PDOP GeoCrd.HDOP GeoCrd.VDOP & Uint16 & Overall precision, horizontal precision and vertical precision; Range: 5 to 999; scaleFactor: -1; offset: 0 & O\\
			\hline
			
			RecCrd.X RecCrd.Y RecCrd.Z & INT16 & Rectangular coordinates X, Y and Z & O\\
			\hline
			
			RecCrd.scaleFactor RecCrd.offset & INT8 INT16 & The scaling factor and offset of cartesian coordinates & O \\
			\hline
			
			RecCrd.PDOP RecCrd.XDOP RecCrd.YDOP RecCrd.ZDOP	& Uint16 & Overall accuracy, XYZ coordinate accuracy, Range:5~999, scaleFactor is -1, offset is 0 & O\\
			\hline
			
		\end{tabular}
		\label{tab1}
	\end{center}
\end{table}

\paragraph{Deleting GetMSVCBValue and SetMSVCBVlaue service}
These two service could set the parameters of IED(electronic transducer) dynamically.However, supporting of them means IED should also deal with MMS message addition to the SV service. This would be difficult task because of the dramatically increased complexity. However, maintaining the two services give the flexibility suitable for the multi-agent system and distributed system, consisting with the trend of smart grid and IOT in the future. A possible solution is to mapping the MMS or ASCI service to SV message. In the early stage, they are delete for simplicity and stability.

\paragraph{Only supporting 80 and 256 points per period} The sample rate and Time Stamp attribute is delete for fixed sampled timestamp. The elements of data-set is also regulated. Considering the limit transmission rate, the reduced SV only support 80 point in one period in early stage, meaning the equivalent sample rate is 4K samples per second(SPS), if the system frequency is supposed to be fixed. A SV message sending time should be less than 1/4K = 250us. Considering the length of SV message and the maximum transmitted bits per seconds of wireless devices, the length should be reduced further.

\begin{figure}[htbp]
	
	\centerline{\includegraphics[width=0.45\textwidth]{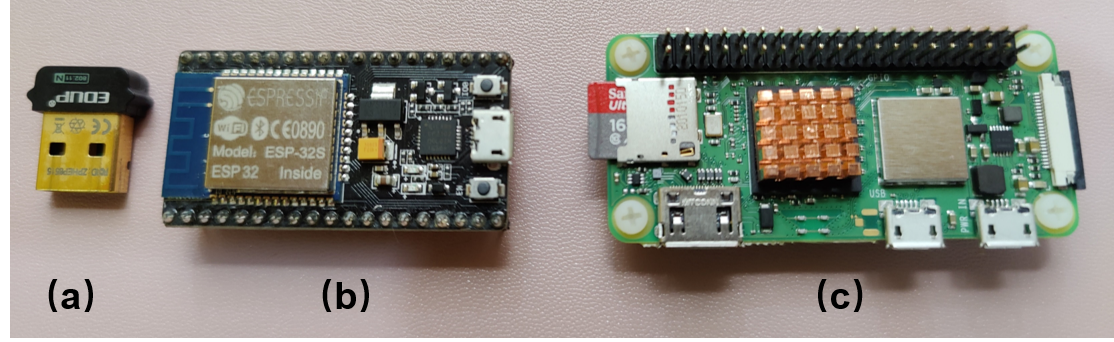}}
	\caption{Wireless IOT Controller.(a) WIFI adapter with USB, provided for gauge.(b) Edge controller, ESP32S.(c)Node controller, Rasbperray Pi zero 2w}
	\label{fig}
\end{figure}

\begin{figure}[htbp]
	
	\centerline{\includegraphics[width=0.4\textwidth]{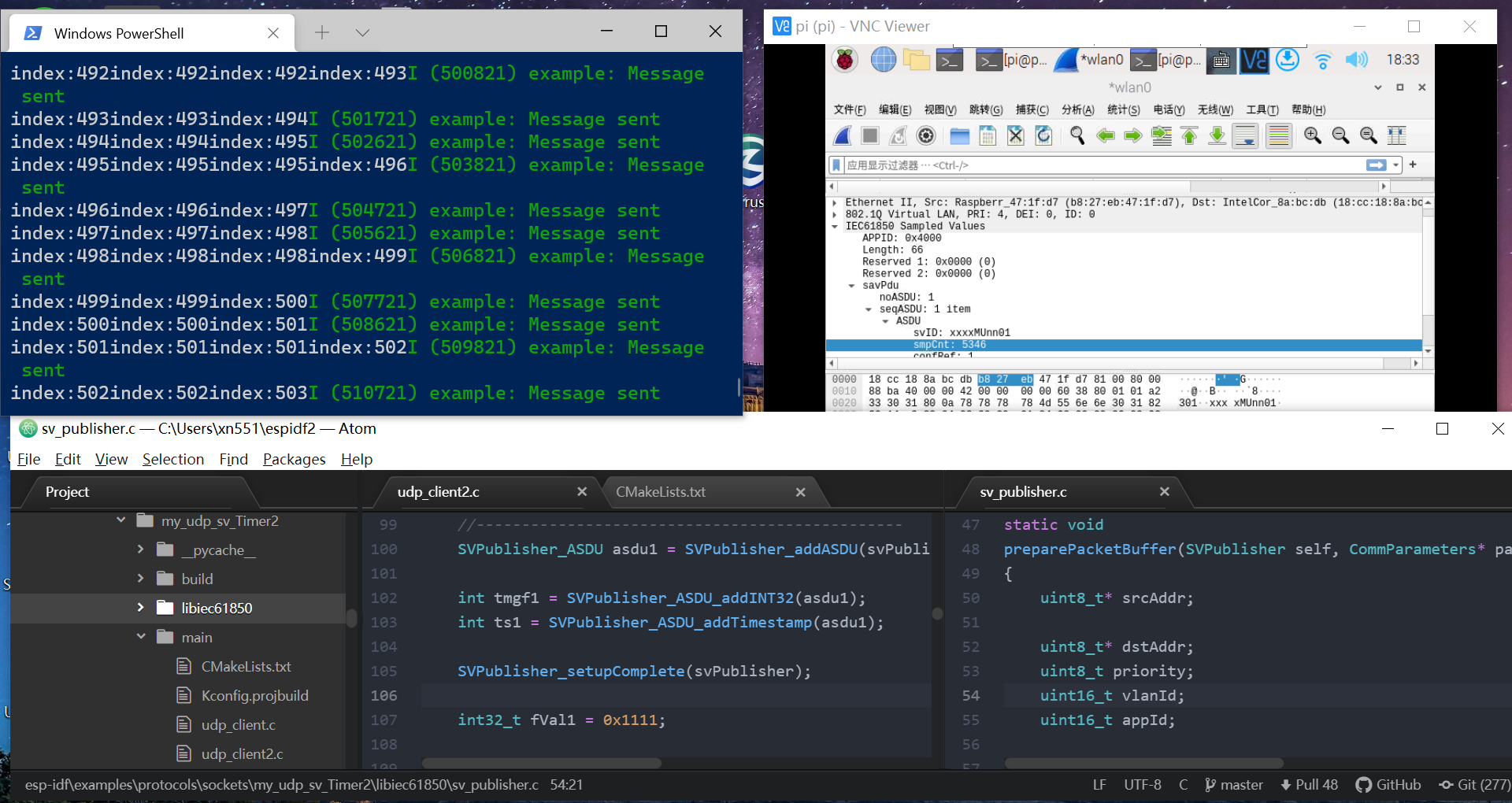}}
	\caption{Develop environment for Reduced IEC61850 SV }
	\label{fig}
\end{figure}

\paragraph{Some attributes is delete or added}
The sample rate and Time Stamp attribute is delete for fixed sampled timestamp.The gain and offset is also fixed, and could be cut. However, for unreliable IOT devices and wireless connection, the timestamp may be reversed for special case. the new added attribute is list in Table III. One class is the absolute geographical coordinates acquired by GPS or BeiDou and their precision, the other is  relative Rectangular Coordinate acquired by local position system. They are necessary especially for cases such as non-contact measurement using quad-rotor unmanned aerial vehicle(UAV).

\subsection{Further reduction considering IOT application}

\paragraph{Adding new Logic Nodes}
A Logic Nodes(LN) TEEF has been added to support the non-contact measurement of electrical field. And There is already a LN for measuring the magnatic field, namely TMGF. TEEF and TMGF has three same data objects, Mode, Beh(Behaviour), Health, Name(Name plate), which are belong to Description data, and they are not transfered by SV message. Only TMGF.MagFld and TEEF.EleFld would be reserved in the dataset and would be sent, they are belong to the SAV common data class(CDC). The LNs supported in SV for Energy IOT are list in Table II.

\paragraph{Redefinition of SAV}
The SAV used for Energy IOT is redefined in Table III. Except for the new added data atrributes to discribe the posiotion information, There are also sereral changes.

\begin{itemize}
\item insMag is redifined as Equation 1 with offset and scaleFactor changed into Int type. 

\begin{equation}
	IntMag = (i+offset)*10^{scaleFactor}
\end{equation}

This is because the Float point type is difficult for MCU to process. The length of offset is also decreased. Since the offset and scaleFactor is fixed for specific data object, they could be omitted for shorter payload.

\begin{table}[htbp]
	\caption{Sample Reduced SV Payload}
	\begin{center}
		\begin{tabular}{|p{0.5cm}|p{1.25cm}|p{1.25cm}|p{4cm}|}
			
			\hline
			\textbf{Bytes No. }&\textbf{Fuction}&\textbf{value } & \textbf{ descriptions}\\
			\hline
			
			6 &	Destination address &	18:cc:18: 8a:bc:db&Mac of host desktop PC\\
			\hline
			
			6	& Source address &	b8:27:eb: 47:1f:d7 & MAC of Raspberr PI \\
			\hline
			
			2	& Type	& 0x81 00 & 802.1Q Virtual Lan\\
			\hline
			
			2 &	PRI, DEI, ID &	0x80 00 & Priority: video, <100ms  DEI: Ineligible; ID:0\\
			\hline
			
			2 &	EtherType &	0x88 ba & IEC 61850/SV \\
			\hline
			
			2 &	APPID &	0x40 00 & \\
			\hline
			
			2 &	Length & 0x00 5c & length: 92\\
			\hline
			
			2&	Reserved 1&	0x00 00& \\
			\hline
			
			2&	Reserved 2&	0x00 00& \\
			\hline
			
			2&	SavPdu &	0x60 38  & Length: 0x52, 82\\
			\hline
			
			3&	noASDU &	0x80 0x01 0x01 & 
			Tag \newline length of length \\
			\hline
			
			2 &	SeqASDU & 0xA2 \newline 0x33 & Tag \newline
			length of SeqASDU, 77 \\
			\hline
			
			2 &	ASDU1 &	0x30 \newline 0x31 & Tag\newline Length of ASDU1, 75 \\
			\hline
			
			11 & svID &	0x80 \newline 0x0a \newline 0x78787878 4d556e3031 & Tag\newline length\newline svID, xxxxMUnn01 \\
			\hline
			
			4 &	smpCnt & 0x82\newline 0x02 \newline 0x0001 & Tag \newline length\newline value 1\\
			\hline
			
			6 &	confRev &	0x83 \newline 0x04 \newline 0x00000001 &Tag \newline length\newline value of confRev\\
			\hline
			
			6 &	refrTm &	0x84 \newline 0x08 \newline 0x00000000 & Tag \newline length \newline  UTC \\
			\hline 
			
			6 &	  smpSynch &  0x85  \newline 0x01  \newline 0x00 & Tag \newline Length \newline none\\
			\hline
			
			14 & seqData &	0x87\newline  0x0e\newline 0x1111 \newline 0x0000 B \newline 0x0000 L \newline 0x0000 H  &Tag \newline  Length \newline TMGF1. MagFld.intMag.i \newline TMGF1. MagFld.GeoCrd.B \newline TMGF1. MagFld.GeoCrd.L \newline TMGF1. MagFld.GeoCrd.H \\
			\hline
			
		\end{tabular}
		\label{tab1}
	\end{center}
\end{table}

\item For quality Q, LE only retained validity and test,  added Derived, which was omitted in this paper and only retained the first two. In IOT application, the q is not Must but optional because of the unreliability of wireless connection.The data reliability needs to be verify by several means, and unreliable data should be discarded. Saving the questioned data will bring in huge workload of data cleaning, especially in the Big-data age. So reserving the data with poor quality is unwise. 

\item The number of ASDU in SavPDU is suggested to be 1. the number of Data attribute in Dataset should not larger than 2. These suggestion grantee the limited length of SV payload.

\end{itemize}

\begin{figure}[htbp]
	
	\centerline{\includegraphics[width=0.4\textwidth]{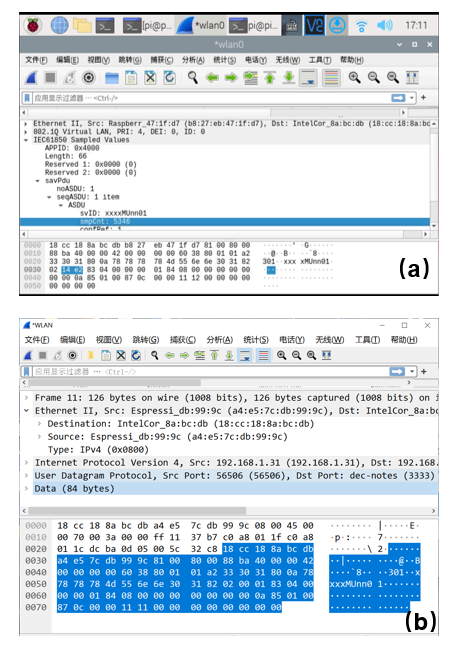}}
	\caption{Analyzing results from Wireshark.(a) SV Payload send by Node controller.(b) SV Payload packed by UDP, send by Edge controller.  }
	\label{fig}
\end{figure}

\section{Implementation}

\subsection{Hardware Specification}
The MCU used for Edge controller is ESP32S developed dy Espressif And Ai-Thinker. ESP32S is a module with Wifi 2.4GHz and Low-power Bluetooth, the chip inside has a dual-core Xtensa LX6 processor with 240MHz frequency, and the Maximum transmission rate of UDP is 30 Mbps for WIFI. The tiny develop-board is shown in Fig 2(b).

The develop-board used for Node is Raspberry Pi zero 2w, as shown in Fig 2(c).To achieve high performance, the Raspberry III or IV could be utilized. Gateway could using MIPS processor for intelligent Router with Open-wrt system. Several chips and develop-board could be selected, for example, Linkit smart 7688 developed by MTK and Seeed.

\subsection{Implement Software}
The main program work is based on ESP32S, the Edge controller. The development enviroment is Espressif IOT Development Framework(IDF), which is based on freeRTOS. The IEC61850-9-2 Sample value service is programmed by transplanting libiec61850, a open-source library of IEC61850. It can run on a Linux system such as Raspberry, Beagle-bone or Linkit smart 7688.

The Program of ESP32S is also based on freeRTOS and LWIP. The libiec61850 \cite{ref5} library is called to generate a SV message template whose elements is described by previous sections. When sample is triggered, the sampled value and time stamp is recorded and packed in SV payload. When the payload is prepared, it gives a message queue to another thread responsible for sending UDP message.

The payload is shown in Table IV, with magnetic field sample value (TMGF1. MagFld.intMag.i) and position information(GeoCrd.B,GeoCrd.L and GeoCrd.H) The total length of payload is 84Bytes. If adding the length of UDP header, the totally length could 126 Bytes(1008 Bits). Choosing 80 samples per period, namely 4K samples per second,  the calculated transmission rate would reach 4.032 Mbps, far less than the Max UDP transmission 30 Mbps.

\subsection{Results}
The debug environment is shown if Fig 3. The program is edit by Atom Editor, and build, flash and monitored in IDF virtual environment running on Power Shell, with USB connection ESP32S develop-board. The Wire-shark program is used to receive and analyze the SV message packed in UDP, as shown in Fig 4(b). 

The desktop also has a VNC viewer to  remote-control the raspberry pi. A IEC61850 SV publisher-client program is used for sending the same raw SV payload without UDP packed, and it is recognized as a 9-2 Sample value message, as shown in Fig 4(a), verifying the reliability of Payload from ESP32S. The sending rate is set to 1 message per second to avoid Network congestion.

\section{Conclusions}
This paper proposed a reduced IEC61850 sample value communication protocols based on UDP to provide a reliable protocols for Energy IOT. The ASDU is largely reduced according to IEC61850-9-2 and LE edition. Some new LN, DO and DA is also expanded to support the IOT application according to IEC61850-7-3 and 7-4. The reduced SV message is also a integrate SV, which could be recognized by wire-shark. The wire-less IOT device such as ESP32S could be utilized for sending the reduced SV message with 4000 samples per seconds.

\begin{IEEEbiography}[{\includegraphics[width=1in,height=1.25in,clip,keepaspectratio]{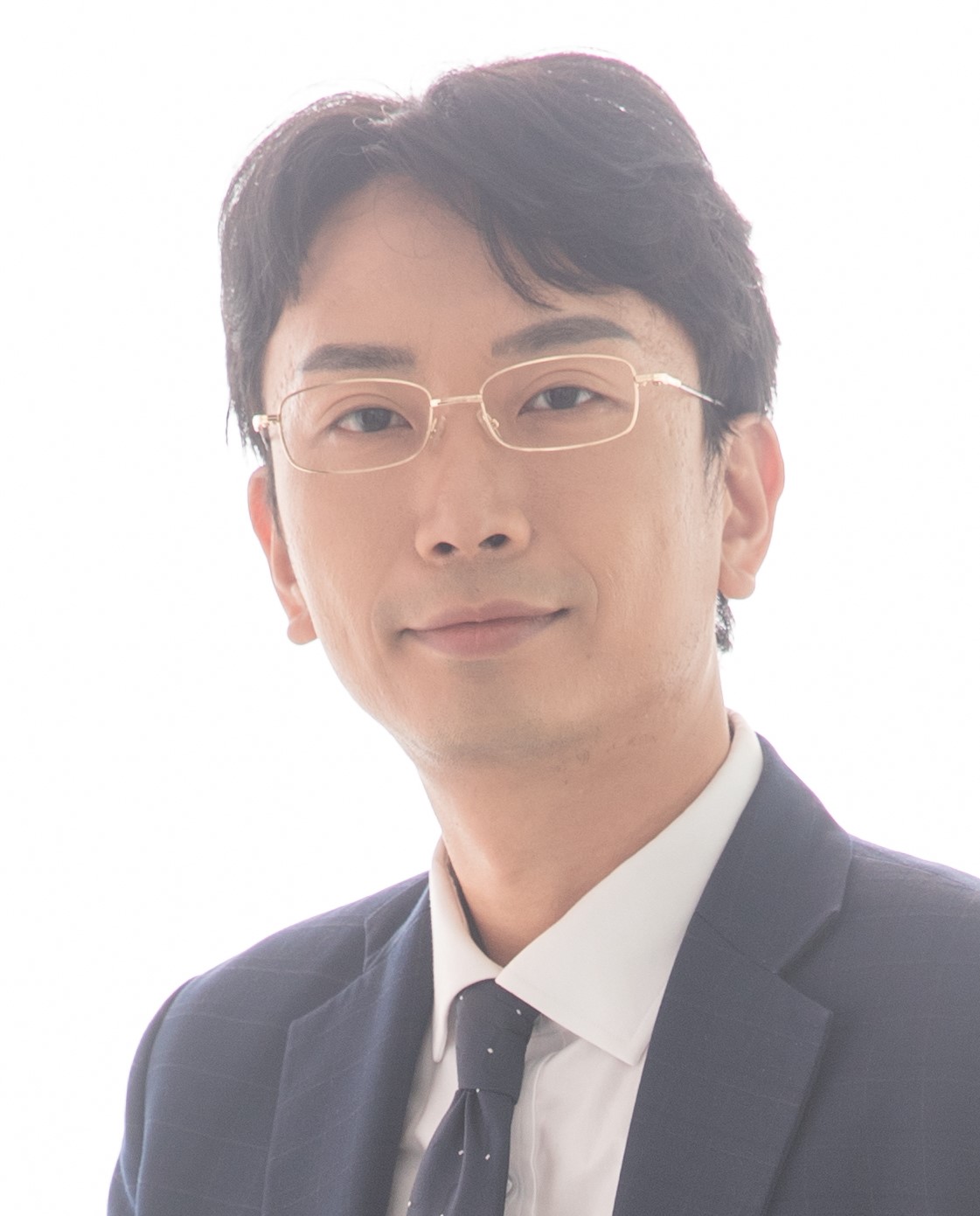}}]{Nan Xie}
	was born in Huainan, China, in 1985. He received the B.S. degree in Material Physics, the M.S. degrees in Material Physics and Chemistry and the Ph.D. degree in Optics from Nankai University, Tianjin, China, in 2007, 2010, and 2013, respectively. He was also co-cultivated at the Institute of Physics, Chinese Academy of Sciences from 2009 to 2013.
	He has been with College of Electrical Engineering and Automation, Fuzhou University since 2013 and is currently working as an Associate Professor since 2021. His research interest includes optical sensor, optical communication,  wireless communication, IEC61850 and Cyber-Physical System, etc.
	
\end{IEEEbiography}

\end{document}